\documentclass{aa}
\usepackage{graphicx}
\begin{document}
\newcommand{\vdag}{(v)^\dagger}
\newcommand{\Msun}{~M_{\odot}}

\title{A Mushroom-shaped Structure from the Impact of a Cloud with the
       Galactic Disk}

\author{Takahiro Kudoh \and Shantanu Basu}
\institute{Department of Physics and Astronomy, University of Western Ontario,
London, Ontario N6A 3K7, Canada}
\offprints{T. Kudoh, 
\email{kudoh@astro.uwo.ca}}
\date{Received / Accepted 3 May 2004}

\abstract{
{We propose that the mushroom-shaped structure of the Galactic worm
GW 123.4--1.5 is created by a cloud collision with the Galactic
gas disk.}
A hydrodynamic simulation shows 
that a mushroom-shaped structure is created after the cloud 
crosses the Galactic midplane. The lifetime of the mushroom-shaped 
structure is of order the dynamical time scale of the disk, 
$\sim 10^7$ years. We find that the velocities across the cap 
of the mushroom-shaped structure in the simulation are consistent 
with the observed values. The simulation also predicts a structure 
on the opposite side of the Galactic plane which is created by 
the Kelvin-Helmholtz instability after the cloud passes 
through the disk.
}

\titlerunning{Mushroom Structure from the Impact of a Cloud}
\authorrunning{Kudoh \& Basu}
\maketitle

%
%________________________________________________________________


\section{Introduction}

The Canadian Galactic Plane Survey (Taylor et al. 2003) revealed 
that the Galactic 
worm candidate GW 123.4--1.5 was an unusual mushroom-shaped cloud
(English et al. 2000). It is hundreds of parsecs in size and
unrelated to conventional shell or chimney structure. 
The mass of the cap is about 4 times greater than that of the stem,
and the total mass is estimated to be about $1.55 \times 10^5 \Msun$,
although such an estimate depends strongly on the distance in which 
there is some ambiguity.
A position-velocity map in the upper portion of the cap shows no 
velocity gradient. However, the lower portion of the two lobes of 
the cap that extend back toward the Galactic plane are blueshifted with 
respect to the central cap region by 5 km/s.

One possible origin for the mushroom-shaped cloud is the rise
of buoyant gas. 
This model was studied by English et al. (2000), in the form
of a supernova event, and  Avillez \& Mac Low (2001),
in the form of bubbles rising from hot gas reservoirs. 
They showed that a mushroom-shaped cloud was created by 
the sweeping effect of the buoyant bubble when the bubble rises from 
the Galactic plane. The gas swept up by the bubble cooled down 
and increased in density. 

In this paper, we propose another possible origin of the 
mushroom-shaped structure: a cloud collision with the Galactic
disk. 
This scenario was first studied by 
Tenorio-Tagle et al. (1986, 1987). However, no one has yet
performed a simulation demonstrating that a cloud impact
would generate the mushroom-shaped structure like GW 123.4--15.

\section{The modeling of a gas collision with the galactic disk}

We perform a hydrodynamic numerical simulation.
As a first step, we simply assume adiabatic gas and axial symmetry. 
We model a local region of the Galaxy and use the axially symmetric 
cylindrical coordinate $(R,z)$, where $R$ represents the distance 
from the axis and $z$ the vertical distance from the midplane 
of the Galaxy.

According to observations, the vertical gravitational field 
of the Galaxy is $g_z = - \alpha z$ near the midplane,
where $\alpha \simeq 2/3 \times 10^{-29} {\rm s}^{-2}$ for the 
solar neighborhood (Spitzer 1978; Bahcall 1984; Tenorio-Tagle et al. 1987).
The hydrostatic equilibrium density of the interstellar isothermal gas 
near the midplane is,
\begin{equation}
\rho(z) = \rho_0 \exp \left[-(z/H_0)^2\right],
\label{eq:gaus}
\end{equation}
where $\rho_0$ is the density at the midplane. 
In equation (\ref{eq:gaus}), the scale length is 
\begin{equation}
H_0=\sqrt{\frac{2}{\gamma\alpha}} c_{s0},
\end{equation}
where $\gamma$ is the specific heat ratio, assumed to be $5/3$, 
and $c_{s0}$ is the sound speed at the midplane.
Since the gravitational field of the Galaxy gradually tends 
to be constant when $z$ is large, we simply let
\begin{equation}
g_z=-\alpha H_g \tanh (z/H_g),
\label{eq:g}
\end{equation}
which tends to $g_z \simeq - \alpha z$ when $z \ll H_g$ and
to $g_z \simeq - \alpha H_g$ when $z \gg H_g$.
We take $H_g=1.5H_0$ in the simulation. Some calculations were done 
by using larger values of $H_g$, but the results were not sensitive 
to it.

We also assume that the interstellar gas has two temperatures 
at the initial time so that
\begin{equation}
T(z)=T_0+0.5(T_1-T_0)\left[1+\tanh\left(\frac{|z|-z_t}{z_d}\right)\right],
\label{eq:T}
\end{equation}
where $T$ is the temperature as a function of $z$.
The temperature at the midplane is the constant $T_0$. 
It becomes another constant $T_1$ at $|z|=z_t$ with 
a transition length of $z_d$.
We take $T_1=10T_0$, $z_t=0.5H_0$ and $z_d=0.1H_0$.
Although observations currently indicate a multi-temperature
galactic atmosphere (e.g., Lockman \& Gehman 1991), 
a two temperature model of the Galactic disk is a first step
that adds a level of sophistication beyond the isothermal
atmospheres explored by Tenorio-Tagle et al. (1986; 1987).

As an initial condition of the Galactic plane, we assume 
a hydrostatic equilibrium of the gas determined by the gravity 
of equation (\ref{eq:g}) and the temperature of equation (\ref{eq:T}).
When $z \ll H_g$, the gravitational field is proportional to $z$ and
the density has a Gaussian distribution like equation (\ref{eq:gaus}). 
There is a transition region of the density at $z_t=0.5H_0$
due to the two temperature model of equation (\ref{eq:T}).
Because the gravity tends to a constant where $z \gg H_g$, 
the density structure finally tends to an exponential atmosphere
at large $z$. The hydrostatic equilibrium of the two temperature 
atmosphere makes a cold dense sheet which is sandwiched between 
hot gas layers. As we will see later, this dense sheet plays 
an important role in making a stem structure of the mushroom.

The density distribution in our simulation is not exactly 
the same as that shown by Dickey \& Lockman (1990) who found 
that the $z$-distribution of the disk density was approximated by 
a summation of two Gaussians which have different velocity 
dispersions. The distribution they found does not show a clear 
discontinuity in the temperature. However, their 
distribution is the average density along the line of sight. 
In this paper, we assume that the cold gas matter is compact
and can be locally well separated from the warm gas.
In addition to this, we assume that the cold gas is located 
in a sheet near the midplane of the disk. 
If the cold gas clouds along the line of sight have
a distribution of heights about the midplane of the Galaxy,
then a transition region of density or temperature would become
unclear in the observations.

We also note that the temperatures ($T_0$ and $T_1$) of the gas
in this simulation are effective temperatures.
Observations show that the scale height of the Galactic 
interstellar medium is much larger than the scale height that 
is estimated from the temperature of the gas. The effective 
temperature that makes the large scale height is considered 
to be caused by turbulent motions. 

Into this equilibrium atmosphere, we introduce a cloud 
with density ($\rho_c$) and $z$-component of velocity ($v_{zc}$):
\begin{equation}
\rho_c(R,z)= \rho_{c0} \left[1- \frac{1}{2}\tanh\left(\frac{r_c(R,z)-r_{wc}}{r_d}\right) \right],
\label{eq:cloud_d}
\end{equation}
\begin{equation}
v_{zc}(R,z) = v_{zc0}  \left[1- \frac{1}{2}\tanh\left(\frac{r_c(R,z)-r_{wv}}{r_d}\right) \right],
\label{eq:cloud_v}
\end{equation}
where
\begin{equation}
r_c(R,z)=\sqrt{R^2+(z-z_c)^2}
\end{equation}
and $z_c$ is the central position of the cloud on the axis.
In equations (\ref{eq:cloud_d}) and (\ref{eq:cloud_v}),
$\rho_{c0}$ and $v_{zc0}$ are constants which parameterize 
the initial cloud density and the $z$-component of the
cloud initial velocity. The parameters $r_{wc}$ and $r_{wv}$ 
determine radii within the cloud for the density and velocity
distributions respectively, and $r_d$ is a transition length.
We use $z_c=-2.5H_0$, $r_{wc}=0.4H_0$, $r_d=0.1H_0$ 
and $r_{wv}=r_{wc}+2r_d$. We let $r_{wv}$ be a little 
larger than $r_{wc}$ in order to avoid a velocity 
gradient within the dense cloud. (If we take $r_{wv}=r_{wc}$, 
a part of the cloud near the edge moves slower than the 
center of the cloud.)

In this simulation, we take $H_0$ as a unit of length, $c_{s0}$ 
as a unit of velocity and $\rho_0$ as a unit of density. 
If we use the value of $\alpha$ for the solar neighborhood and 
$c_{s0} \simeq 10$ km/s (which corresponds to $T_0 
\sim 10^4$ K) as the turbulent velocity in the Galaxy,
then $H_0 \simeq 140$ pc. The unit of time is 
$t_0=H_0/c_{s0} \simeq 1.4 \times 10^7$ year.
The important parameters in this simulation are the initial density
of the cloud, $\rho_{c0}$, and the initial velocity of the cloud, 
$v_{zc0}$. In this paper, we will show the result of $\rho_{c0}=\rho_0$,
which means that the initial density of the cloud is the same as 
that of the cold dense sheet on the midplane of the Galaxy
(e.g., $\sim 1$ cm$^{-3}$). We will show two cases of the initial 
velocity; $v_{zc0}=10c_{s0} \simeq 100$ km/s and
$v_{zc0}=5c_{s0} \simeq 50$ km/s.
The former is the order of the line-of-site motion of a high-velocity 
cloud (Wakker \& van Woerden 1997), and 
the latter corresponds to a intermediate-velocity cloud 
(Danly 1989; Kuntz \& Danly 1996).

We use the CIP method (e.g., Yabe, Xiao \& Utsumi 2001) to solve
the finite difference equations of hydrodynamics. 
The minimum grid size is $0.01H_0$ in both the $R$ and $z$ 
directions. In the $R$-direction, we gradually increase the grid size 
where $R > 3H_0$. The computational region is $0 \leq R \leq 10H_0$, 
$-5H_0 \leq z \leq 10H_0$. The outer boundary conditions are 
free boundaries for both the $R$ and $z$ directions. 

\section{Results}

\subsection{The case of $v_{zc0}=10 c_{s0} \simeq 100$ km/s}

\begin{figure*}
\centering
\includegraphics[width=16cm]{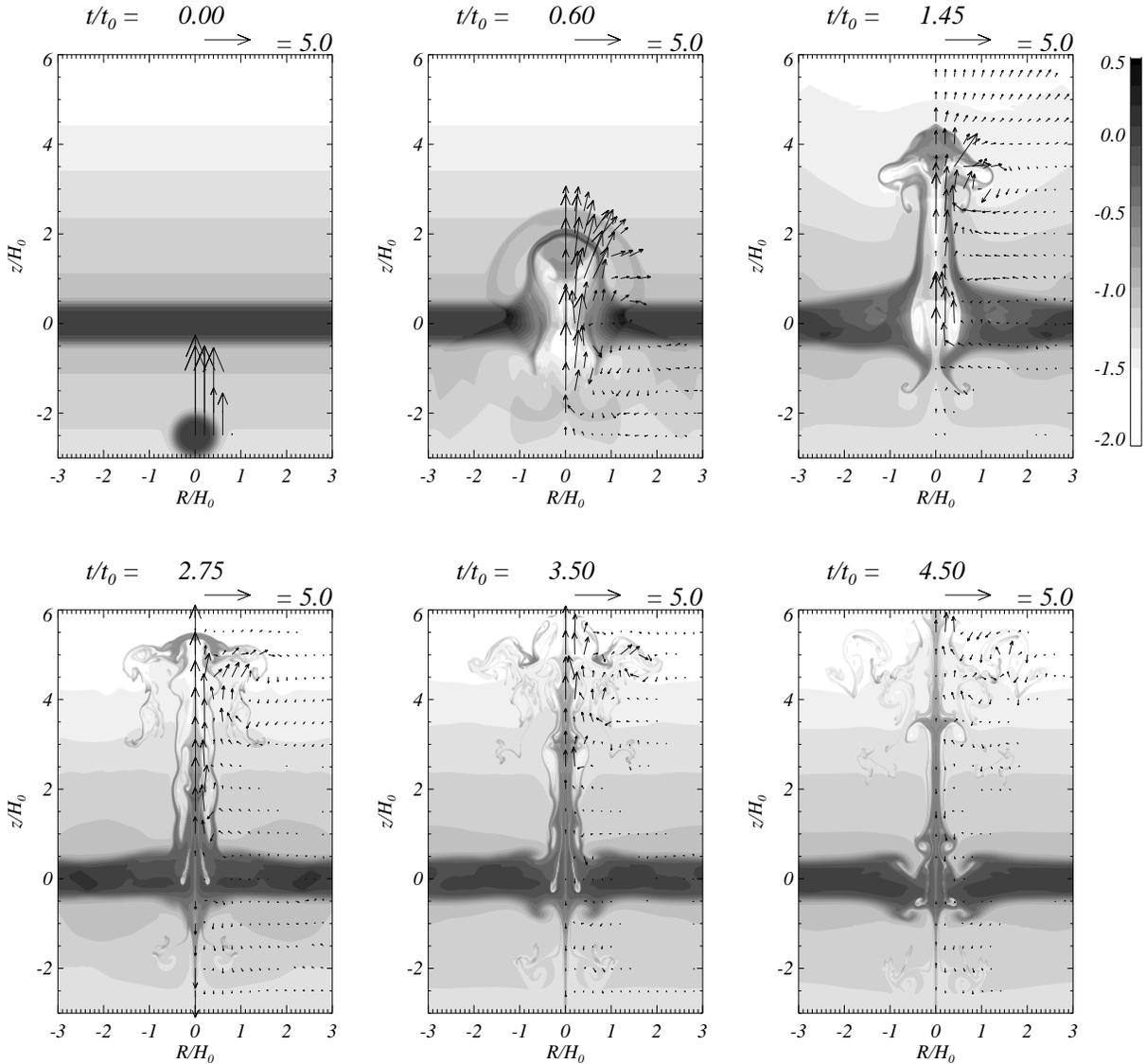}
\caption{Time evolution of density in the case of $v_{zc0}=10 c_{s0} 
\simeq 100$ km/s . The gray scale shows the logarithmic scale of the 
density which is normalized by the initial density at $z=0$, $\rho_0$. 
Arrows represent the velocity vectors which are normalized by the initial 
sound speed at $z=0$, $c_{s0}$. Velocity vectors are shown only 
in the region $R \ge 0$.}
\end{figure*}

Figure 1 shows the time evolution of the density.
The top-left panel shows the initial condition ($t=0$).
The initial atmosphere is hydrostatic equilibrium determined by
equations (\ref{eq:g}) and (\ref{eq:T}). A gas cloud is input 
in the atmosphere according to the equations 
(\ref{eq:cloud_d}) and (\ref{eq:cloud_v}).

As the cloud impacts the midplane of the Galaxy, it goes through
the dense sheet in the Galactic plane ($t=0.6t_0$). After that, 
the cloud is decelerated by gravity ($t=1.45t_0$) and 
is broken by 
{eddies of the hot gas flow generated behind the
cloud after it passes through the disk ($t>2.75t_0$).}
The cloud broken by eddies forms a cap of a mushroom-shaped 
structure.

When the gas cloud passes though the Galactic plane, a part of 
the dense sheet in the Galactic plane goes along with the cloud 
and forms a shell-like structure ($t=0.6t_0$). 
This shell is elongated and forms the stem of the
mushroom-shaped structure ($t=1.45t_0 - 2.75t_0$). 
Moreover, when the hole that was made by the cloud passage
is refilled with the gas on the plane, a flow appears near the 
axis ($t=1.45t_0 - 2.75t_0$). This flow lifts a part of the dense 
gas in the Galactic plane, which also contributes to the 
stem ($t=3.5t_0$). 
{These processes show the importance of the dense sheet 
in the Galactic atmosphere that results from our assumed transition 
in effective temperature.}
During this process, the stem is getting narrow 
($t=2.75t_0 - 4.5t_0$).
The simulation also shows the development of small structures 
on the opposite side of the Galactic plane ($z<0$). 
The two-sided eddies, evident at $R=\pm H_0$ and $z=-1.5H_0$ 
in Figure 1 at $t=2.75t_0$, are created by the Kelvin-Helmholtz 
instability after the cloud passes through the disk ($t=1.45t_0$). 
Finally, the cap is diffused by eddies and becomes unclear 
($t>4.5t_0$). The lifetime of the mushroom-shaped 
structure is the order of the dynamical time scale, 
$t \sim t_0 \sim 10^7$ year.
The movie for the density clearly shows the time evolution.

\begin{figure}
\resizebox{\hsize}{!}{\includegraphics{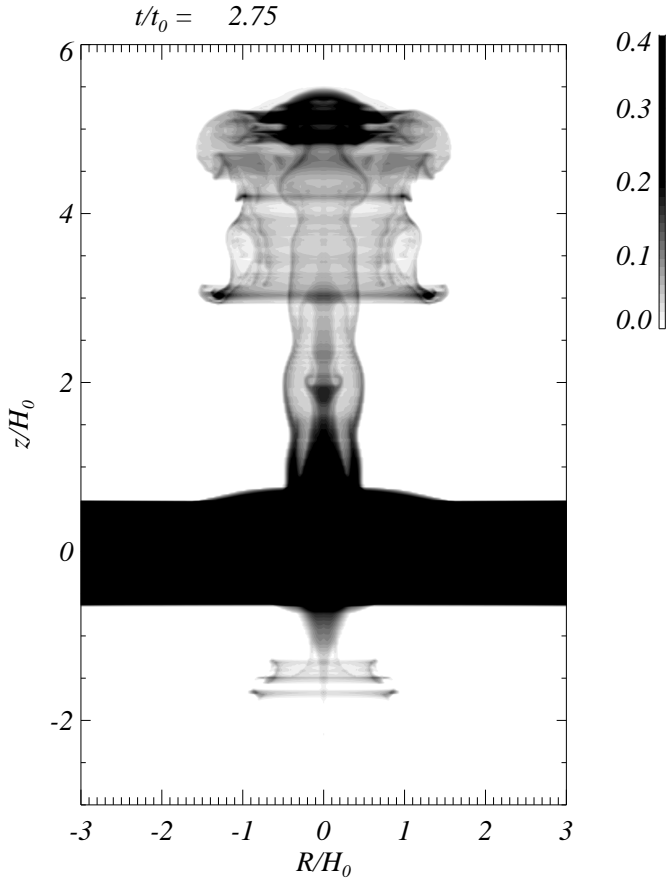}}
%\resizebox{\hsize}{\includegraphics{f2.eps}}
\caption{
Column density of low temperature gas ($T < 0.8 T_1$) at $t=2.75t_0$
in the case of $v_{zc0}=10 c_{s0} \simeq 100$ km/s.
The gray scale shows the linear scale of the column density
normalized by $\rho_0 H_0$. 
}
\end{figure}

Figure 2 shows a snapshot of the column density at $t=2.75t_0$. 
Since this is an axially symmetric simulation, 
we can integrate the density perpendicular to the $R-z$ plane 
to get a column density. Here, we have integrated 
the low temperature gas ($T<0.8T_1$) when we calculate 
the column density, in order to avoid including the gas 
which is surrounding the mushroom-shaped structure. 
Sampling the low temperature gas makes the mushroom-shaped 
structure clear to understand,
though this is not exactly the same column density 
which is obtained by observations.
The column density of low temperature gas
has a clear stem and cap structure which resembles
the mushroom-shaped structure of GW 123.4--15.
(The movie for the column density also shows 
its time evolution.) The width ratio of the cap and stem 
at this snapshot is about 3:1, and the mass ratio is about 2.4:1. 
These results are consistent with a observational ratio of 3:1, 
and the mass ratio of 4:1. However, the maximum height of 
the mushroom at $t=2.75t_0$ is about $5.5H_0 \simeq 770$ pc, 
which is greater than the mushroom-shaped structure 
of GW 123.4--15 though the scale depends on the 
ambiguous distance.

The column density in Figure 2 also shows 
a smaller structure on the opposite side of the Galactic plane 
(around $z=-1.5H_0$). This structure is created by 
the Kelvin-Helmholtz instability after the cloud passes 
through the disk. The lifetime of this structure is 
also $\sim 10^7$ year.

\begin{figure}
\resizebox{\hsize}{!}{\includegraphics{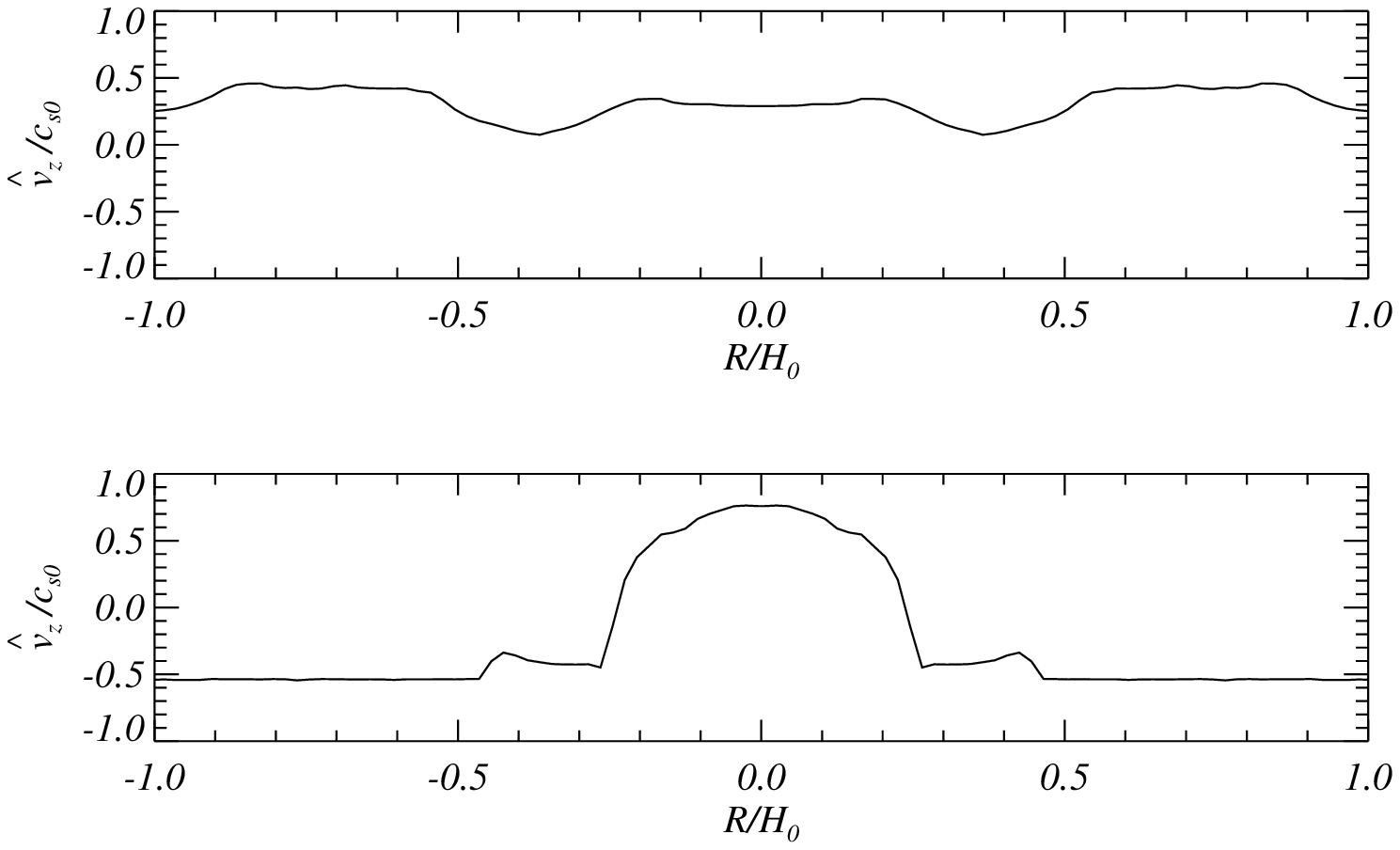}}
\caption{
Averaged velocity across the cap as a function of $R$
in the case of $v_{zc0}=10 c_{s0} \simeq 100$ km/s. 
The upper panel shows the velocity along $z=5.0H_0$ and the bottom 
panel along $z=3.0H_0$.
}
\end{figure}

Another important comparison with observations is the velocity
structure of the mushroom-shaped structure. 
The observational velocity, which is obtained by 
the Doppler shifts of neutral hydrogen, 
{is considered to be sampling a component of $z$-velocity}
if the mushroom-shaped structure is tilted
relative to the line of sight.
In order to show the velocity structure from our simulation, 
we calculate the averaged $z$-component of the velocity 
along a line of sight perpendicular to the $R-z$ plane. 
The averaged velocity is 
calculated in the following manner:
\begin{equation}
\hat{v}_z=\frac{ \int \rho v_z dl}{\int \rho dl} ,
\end{equation}
where $l$ is the length along a line perpendicular to 
the $R-z$ plane. The integration was also done only for 
the cold gas in the same way that we get the column density.
Figure 3 shows the averaged velocity across the cap.
The top panel shows the velocity along $z=5.0H_0$, 
and the bottom along $z=3.0H_0$. 
At the lower portion of the cap ($z=3.0H_0$), the flow goes 
up near the axis but goes down near the edge of the cap, 
while all of the cap is gradually going up in the 
upper portion ($z=5.0H_0$). The velocity of the 
upper portion ($\hat{v}_z \simeq 0.4c_s$) is in between 
the maximum and minimum velocities of the lower portion of the cap.
These velocity structures are consistent with 
Figures 1c and 1d of English et al. (2000).

\subsection{The case of $v_{zc0}=5 c_{s0} \simeq 50$ km/s}

\begin{figure*}
\centering
\includegraphics[width=16cm]{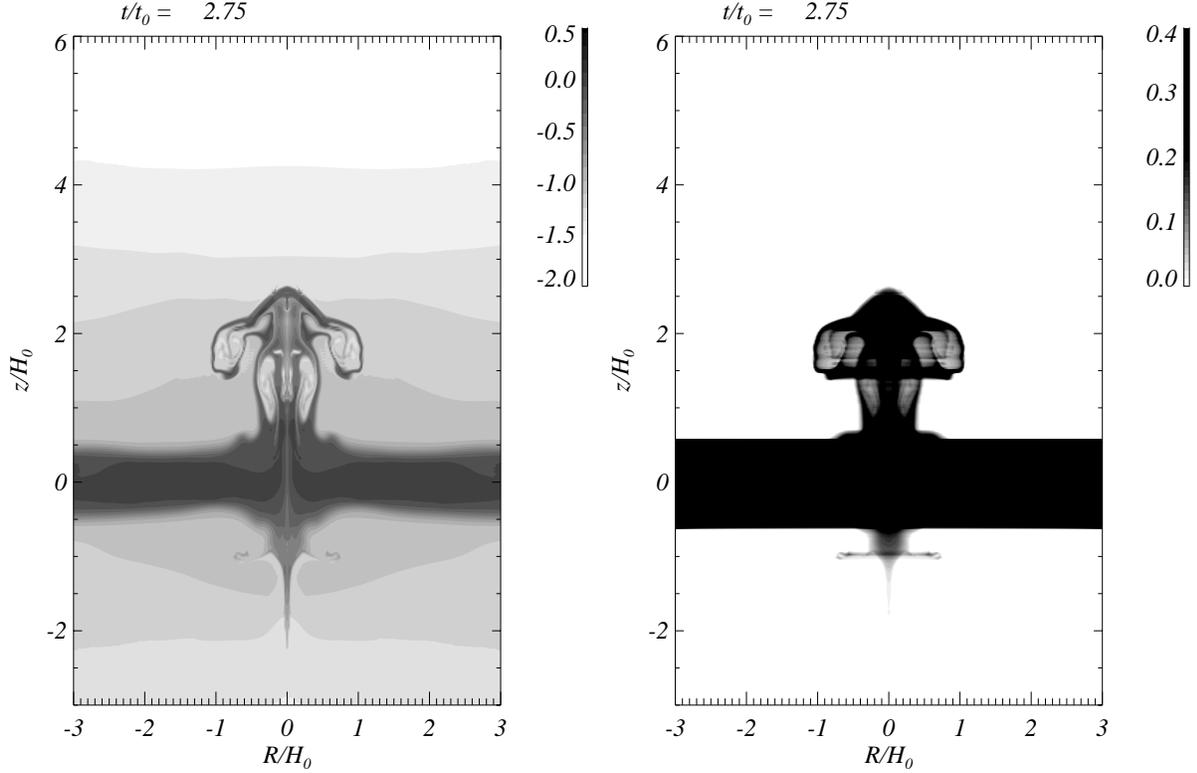}
\caption{
Results in the case of $v_{zc0}=5 c_{s0} \simeq 50$ km/s.
Left: density at $t=2.75t_0$. Right: column density of 
low temperature gas ($T < 0.8 T_1$) at $t=2.75t_0$.
The density is normalized by $\rho_0$ the initial density at $z=0$. 
The column density is normalized by $\rho_0 H_0$. 
}
\end{figure*}

Here, we will show the case of $v_{zc0} 
= 5c_{s0} \simeq 50$ km/s. The time evolution of the cloud 
is similar to the case of $v_{zc0} = 10c_{s0} \simeq 100$ km/s. 
Figure 4 shows the snapshot of density and column density at
$t=2.75t_0$.
We can see the similar mushroom-shaped structure around 
$t=2.75t_0$, though the size of the mushroom is a little 
smaller than the case of $v_{zc0} = 10c_{s0} \simeq 100$ km/s.
The maximum height of the mushroom at $t=2.75t_0$ is
about $3H_0 \simeq 420$ pc, which is almost the same
size as that of GW 123.4--15. The width ratio of the cap
and stem is about 2:1 and the mass ratio is about 4:1, 
which also show good agreements of the observation.

Figure 5 shows that averaged velocity across the cap
at $t=2.75t_0$.
The top panel shows the velocity along $z=2.0H_0$, 
and the bottom along $z=1.4H_0$. 
The lower portion of the cap shows that the flow goes 
up near the axis but goes down near the edge of the cap, 
while the upper portion of the cap shows the flow 
goes up in wider region than that of the lower cap.
This is also consistent with the observation.
Movies of the density and column density for
this model further illustrate the evolution.

\subsection{Parameter study}

We have performed a parameter study for the velocity ($v_{zc0}$) 
and density ($\rho_{c0}$) of the impact cloud, as well as the
specific heat ratio ($\gamma$).

When the velocity is $3c_{s0}$ ($\simeq 30$ km/s),
the mushroom-shaped structure does not appear.
The cloud does not reach enough height after it goes
though the dense sheet, so that the cap and stem 
structure is not apparent (the cap is prominent, 
but the stem is not). Therefore, we need at least 
about 50 km/s to get a mushroom-shaped structure when 
the density of the cloud is the same order of that
of the dense sheet on the Galactic plane.

If the density of the cloud is much less than that
of the Galactic plane ($\rho_c < 0.1\rho_0$), 
our simulations show that the cloud does not go through 
the plane even in the case of $v_{zc0}=10c_{s0} \simeq 100$ km/s.
Therefore, the cloud density should be of the same order 
as that of the dense sheet of the Galactic plane 
unless the velocity of the cloud is extremely high. 

In order to see the effect of cooling mechanisms
(radiative cooling and/or dissipation of turbulence),
we ran simulations with $\gamma=1.05$. Even for
this small specific heat ratio, we had similar
mushroom-shaped structures both for high and low
velocity cases. Therefore, we think that the effect 
of cooling does not affect the basic result.
However, we found  some differences: (1) the impact cloud 
reaches a higher position (about 20\% higher) than that of 
the adiabatic case after it goes through the dense sheet. 
(2) Both stem and cap become a little narrower 
(by about 10\%) than those of the adiabatic case. 

\begin{figure}
\resizebox{\hsize}{!}{\includegraphics{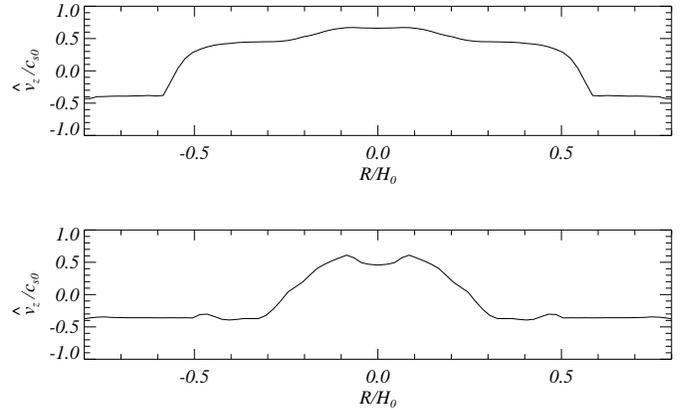}}
\caption{
Averaged velocity across the cap as a function of $R$
in the case of $v_{zc0}=5 c_{s0} \simeq 50$ km/s. 
The upper panel shows the velocity along $z=2.0H_0$ and the bottom 
panel along $z=1.4H_0$.
}
\end{figure}

\section{Discussion and Conclusion}

Based on our simulations, we conclude that a cloud collision 
with the Galactic disk is one of the possible models of 
the mushroom-shaped structure, GW 123.4--1.5. 
The density and column density clearly show a cap and
stem structure which resembles the observed structure
for both cases of a high-velocity cloud 
($v_{zc0}=10 c_{s0} \simeq 100$ km/s) and 
an intermediate-velocity cloud ($v_{zc0}=5 c_{s0} \simeq 50$ km/s).
The case of $v_{zc0}=5 c_{s0} \simeq 50$ km/s shows 
a better agreement with the size of the mushroom-shaped structure
as well as the mass ratio of the cap and stem. 
It may mean that GW 123.4--1.5 is created by the collision of 
an intermediate-velocity cloud into the Galactic plane,
though the stem in our simulation is a little thicker than 
that of the observation.

However, mushroom-shaped structures may not be commonly observed
unless the densities of the impact cloud are comparable to the 
density of the Galactic plane and the frequency of the impacts 
is high. In order to get a mushroom-shaped 
structure, we found that the cloud density should be at least 
the same order as the disk density. Given a disk number 
density $\sim 1$ cm$^{-3}$, the column density of the impact cloud 
is $> 10^{20}$ cm$^{-2}$. Observations show that such large 
column densities are not common both for high-velocity clouds 
(Wakker \& von Woerden 1997) and intermediate-velocity clouds 
(Benjamin \& Danly 1997). Nevertheless, there are some exceptions, 
such as the Ursa Major intermediate-velocity cloud
(Snowden et al. 1994; Benjamin et al. 1996)
which has velocity $\sim 50$ km/s,
column density $\sim 2.0 \times 10^{20}$ cm$^{-2}$, and estimated
size ($\sim 15 \times 50$ pc) which implies a number density
$\sim 1$ cm$^{-3}$.
The numerical simulations also show that the lifetime of 
the mushroom-shaped structure is only about a dynamical time, 
$\sim 10^7$ years. These results may be some of the reasons 
that only one mushroom-shaped structure has yet been found 
near the Galactic plane.

The cloud collision model in this paper is an alternative
to the buoyant model. The velocity structures across the cap 
in this model are 
consistent with observations of GW 123.4--1.5, 
but it is not obvious that the buoyant model has 
a similar velocity structure. 
More detailed kinematical analysis of 
GW 123.4--1.5 would distinguish the two models. 
In addition to the velocity, the stem near the midplane 
of the Galaxy is clear in our model, while it is not clear 
in the buoyant model. Our model shows that the stem directly 
connects to the dense gas near the midplane of the Galaxy, 
which seems to be the same as the observation. 
Moreover, our model predicts 
small structures on the opposite side of the Galactic plane 
(in Figure 2 around $z=-1.5H_0$, and in Figure 4 around $z=-1.0H_0$). 
These structures may not be significant in current observations, 
but may be an interesting target in the future in order to distinguish 
the two models.

{ 
We have assumed a Galactic atmosphere with two effective temperatures.
The resulting dense sheet of gas near the midplane
contributes to the formation of the stem of the mushroom-shaped
structure. In our model, the stem originates from material
lifted up from the dense sheet, while the cap
is formed primarily by the breakup of the impact cloud.
A dense stem is actually not clear in
single-temperature models that we performed with
the same initial impact clouds. A distinct transition
between cold (weakly turbulent) and warm (highly turbulent) gases is not
obvious from observations which measure the integrated gas along the
line of sight. However, the mushroom-shaped structure we have
obtained in our simulations may suggest that the cold gas clouds 
which are located near the midplane of the Galaxy are locally well 
separated from the warm gas.
}

Future work can improve upon this model by including a 
self-consistent treatment of turbulent pressure, and
the effect of radiative cooling of the thermal gas.
Moreover, the magnetic pressure 
is comparable to the effective turbulent pressure 
in the ISM, and a collision exactly 
perpendicular to the Galactic plane would be rare in 
a real situation. Three-dimensional simulations including 
the magnetic field, heating and cooling, and turbulent gas 
would be an ultimate work for the future in order to have more
detailed comparison between the numerical simulations
and observations.

\begin{acknowledgements}

The authors acknowledge comments from and discussions with 
Jayanne English and Ashish Asgekar. 
{The authors also wish to thank the anonymous referee 
for comments which substantially improved the paper.} 
TK acknowledges support due to a Fellowship from the Canadian 
Galactic Plane Survey. TK also benefited from a Fellowship 
from SHARCNET, a high-performance computing project funded 
by ORDCF and CFI/OIT. 
SB was supported by an individual research grant from NSERC. 
Numerical computations were carried out {mainly} on 
the VPP5000 at the Astronomical Data Analysis Center in 
the National Astronomical Observatory of Japan.

\end{acknowledgements}

\end{document}